\documentclass{PoS}

\usepackage[numbers]{natbib}
\usepackage{url}
\usepackage{threeparttable}

\title{On the radio and GeV-TeV $\gamma$-ray emission connection in \fermi\ blazars.}

\ShortTitle{Radio VLBI and GeV-TeV $\gamma$-ray emission connection.}

       \author{\speaker{R.~Lico}$^{,1,2}$, M.~Giroletti$^{1}$, M.~Orienti$^{1,2}$, L.~Costamante$^{3}$,  V.~Pavlidou$^{4}$, F.~D'Ammando$^{1,2}$, F.~Tavecchio$^{5}$. \\

$^{1}$INAF Istituto di Radioastronomia, Bologna, Italy. \\
$^{2}$Dipartimento di Fisica e Astronomia, Universit\`a di Bologna, Italy. \\
$^{3}$ASI - Unit\`a Ricerca Scientifica, Roma, Italy. \\
$^{4}$Department of Physics and Institute for Plasma Physics, University of Crete, Greece. \\
$^{5}$INAF - Osservatorio Astronomico di Brera, Merate, Italy. \\

        E-mail: \email{rocco.lico@unibo.it}
}

\abstract{The \fermi -LAT revealed that the census of the $\gamma$-ray sky is dominated by blazars. Looking for a possible connection between radio and $\gamma$-ray emission is a central issue for understanding the blazar physics, and various works were dedicated to this topic. However, while a strong and significant correlation was found between radio and $\gamma$-ray emission in the 0.1-100\,GeV energy range, the connection between radio and very high energy (VHE, E>0.1 TeV) emission is still elusive. The main reason is the lack of a homogeneous VHE sky coverage, due to the operational mode of the imaging atmospheric Cherenkov telescopes.
With the present work we aim to quantify and assess the significance of the possible connection between high-resolution
radio emission, on milliarcsecond scale, and GeV-TeV $\gamma$-ray emission in blazars. For achieving our goal we extract two large and unbiased blazar samples from the 1FHL and 2FHL \fermi\ catalogs, above 10\,GeV and 50\,GeV, respectively. 
To investigate how the correlation evolves as the $\gamma$-ray energy increases, we perform the same analysis by using the 0.1-300\,GeV 3FGL $\gamma$-ray energy fluxes.
When we consider the 0.1-300\,GeV $\gamma$-ray energy range, we find a strong and significant correlation for all of the blazar sub-classes. Conversely, when we consider the $\gamma$-ray emission above 10\,GeV the correlation with the radio emission vanishes, with the exception of the blazar sub-class of high synchrotron peaked objects.
}

\FullConference{7th Fermi Symposium 2017\\
		15-20 October 2017\\
		Garmisch-Partenkirchen, Germany}

\newcommand{\fermi}{\textit{Fermi}} 

\begin{document}

\section{Introduction and scientific context}
Blazar objects belong to the family of radio-loud active galactic nuclei (AGNs), with their relativistic jets closely aligned to the observer's line of sight. 
In general, according to the presence or not of broad emission lines in their optical spectra, blazars are commonly classified as flat spectrum radio quasars (FSRQs) or BL Lac objects (BL Lacs) \citep[e.g.,][]{Stickel1991}. 
A further classification of blazars is based on the position of peak of the low-frequency synchrotron component ($S_{\mathrm{Syn}}^{\mathrm{peak}}$) in theirs spectral energy distribution (SED) \citep{Abdo2010}. 
When $S_{\mathrm{Syn}}^{\mathrm{peak}}$ is found at frequencies $<10^{14}$\,Hz blazars are classified as low-synchrotron peaked (LSP), while when $S_{\mathrm{Syn}}^{\mathrm{peak}}$ is at frequencies $>10^{15}$\,Hz they are classified as high-synchrotron peaked blazars (HSP); in the intermediate case, they are classified as intermediate synchrotron peaked (ISP).

When the \fermi\ large area telescope (LAT) confirmed that blazars dominate the census of the $\gamma$-ray sky \citep{Acero2015}, the possible connection between radio and $\gamma$-ray emission, which is essential to understand the blazar physics and the emission processes, aroused a great interest \citep[e.g.\,,][]{Kovalev2009, Ghirlanda2010, Giroletti2010, Piner2014, Giroletti2016}.
In \citep{Ackermann2011} a strong and significant correlation between radio emission (by using 8\,GHz interferometric and 15\,GHz single dish data) and $\gamma$-ray emission in the 0.1-100\,GeV energy range, was found for a large AGN sample. In \citep{Ackermann2011} it emerged that the correlation strength depends on the simultaneity, on the blazar type and on the energy band: in particular, it was found that the correlation weakens when higher $\gamma$-ray energies are considered.

With the present work we aim to investigate how the radio and $\gamma$-ray emission correlation evolves when the very high energy (VHE, $E > 0.1$\,TeV) $\gamma$-ray band is considered. 
The possible connection between radio and VHE $\gamma$-ray emission is still elusive, due to the lack of a homogeneous coverage of the VHE sky. Because of the operational mode of the imaging atmospheric Cherenkov telescopes, mainly operating in pointing mode and with a limited field of view, the current VHE catalogs are strongly biased. In this context the new generation Cherenkov telescope array \citep[CTA, ][]{Actis2011} will be a valuable resource.
For our goal we use the first and second \fermi -LAT catalogs of high-energy $\gamma$-ray sources, 1FHL \citep[][]{Ackermann2013} and 2FHL \citep[][]{Ackermann2016}, which provide us with two large and unbiased samples of $\gamma$-ray sources detected above 10\,GeV and 50\,GeV, respectively. At radio frequencies we make use of high-resolution very long baseline interferometry (VLBI) observations, which are more suitable for probing the source innermost regions, where $\gamma$-rays are supposed to be produced.

\section{Correlation results}
The 1FHL (10-500\,GeV energy range) and 2 FHL (50\,GeV - 2\,TeV energy range) are based on $\gamma$-ray data obtained during the first 3 and 6.5 years of the \fermi\ mission, respectively. In both catalogs $\sim80$\% of the sources are AGNs and the vast majority of them are classified as blazars. 

By focusing on the $\gamma$-ray sources with declination $\delta>0^\circ$, for our analysis we select a sample of 237 1FHL (hereafter 1FHL-n sample) and 131 2FHL (hereafter 2FHL-n sample) sources. At radio frequencies we use the VLBI flux densities (at 5 or 8\,GHz) provided by the radio fundamental catalog\footnote{\url{http://astrogeo.org/rfc/}.} (RFC). For further details about the sample selection and construction see \citep{Lico2017}.
To investigate how the correlation between radio end $\gamma$-ray emission evolves as higher $\gamma$-ray energies are considered, we perform the same correlation analysis by using the third \fermi -LAT source catalog \citep[3FGL, ][]{Acero2015} $\gamma$-ray energy fluxes, in the 0.1 - 300\,GeV energy range.

To asses the correlation statistical significance we use the method proposed by \citep{Pavlidou2012}, which is based on the permutations of luminosities, and takes into account the various observational biases and distance effects, which could affect the correlation results. This method was also used in the correlation analysis presented in \citep{Ackermann2011}.

\subsection{1FHL-n and 2FHL-n samples}
\label{corr_results}

\begin{figure*}[t]
\begin{center}
\includegraphics[bb=6 0 495 350, width= 0.45\textwidth, clip]{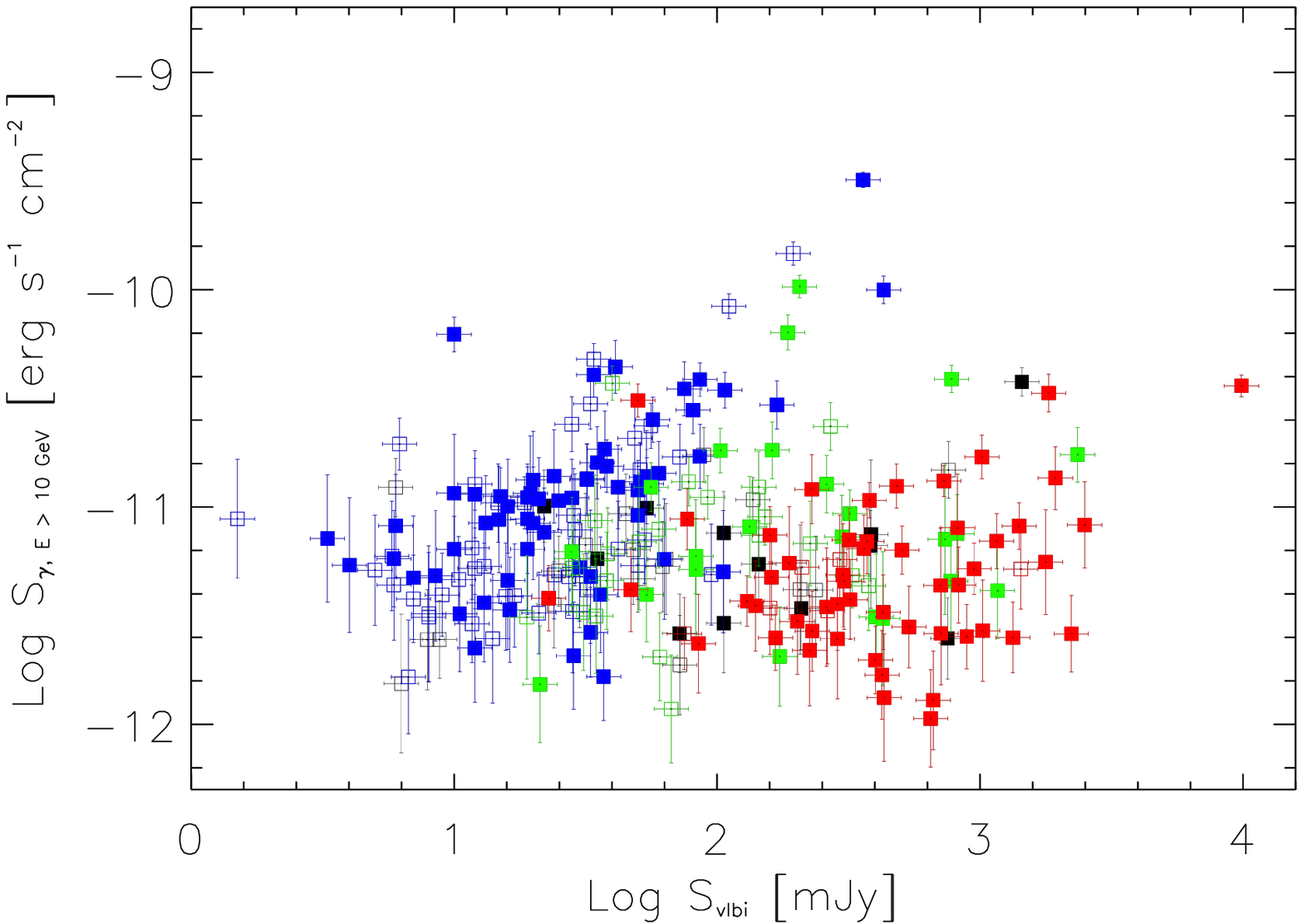} 
\includegraphics[bb=0 0 485 350, width= 0.45\textwidth, clip]{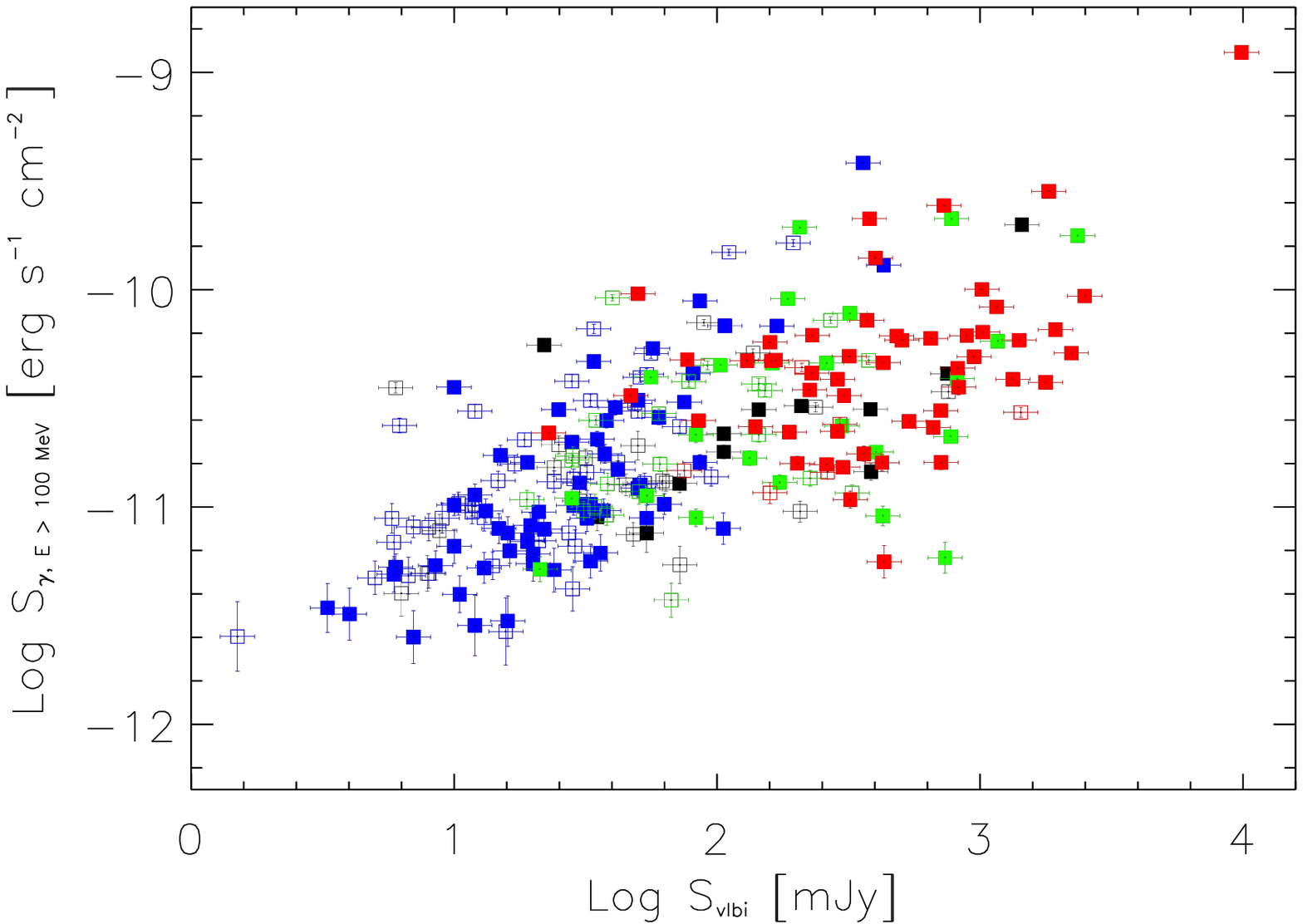}  \\
\includegraphics[bb=6 0 495 348, width= 0.45\textwidth, clip]{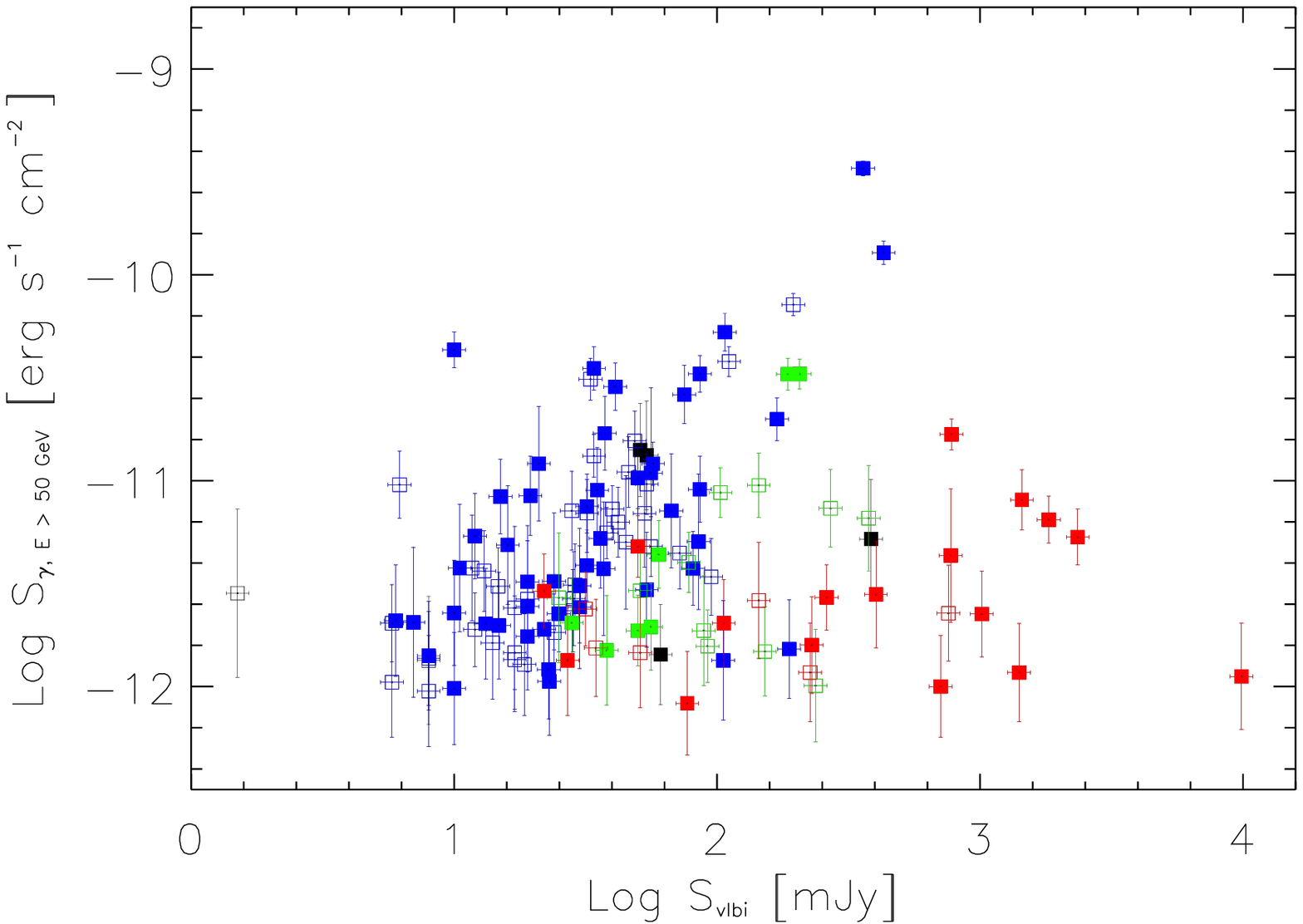} 
\includegraphics[bb=0 0 485 348, width= 0.45\textwidth, clip]{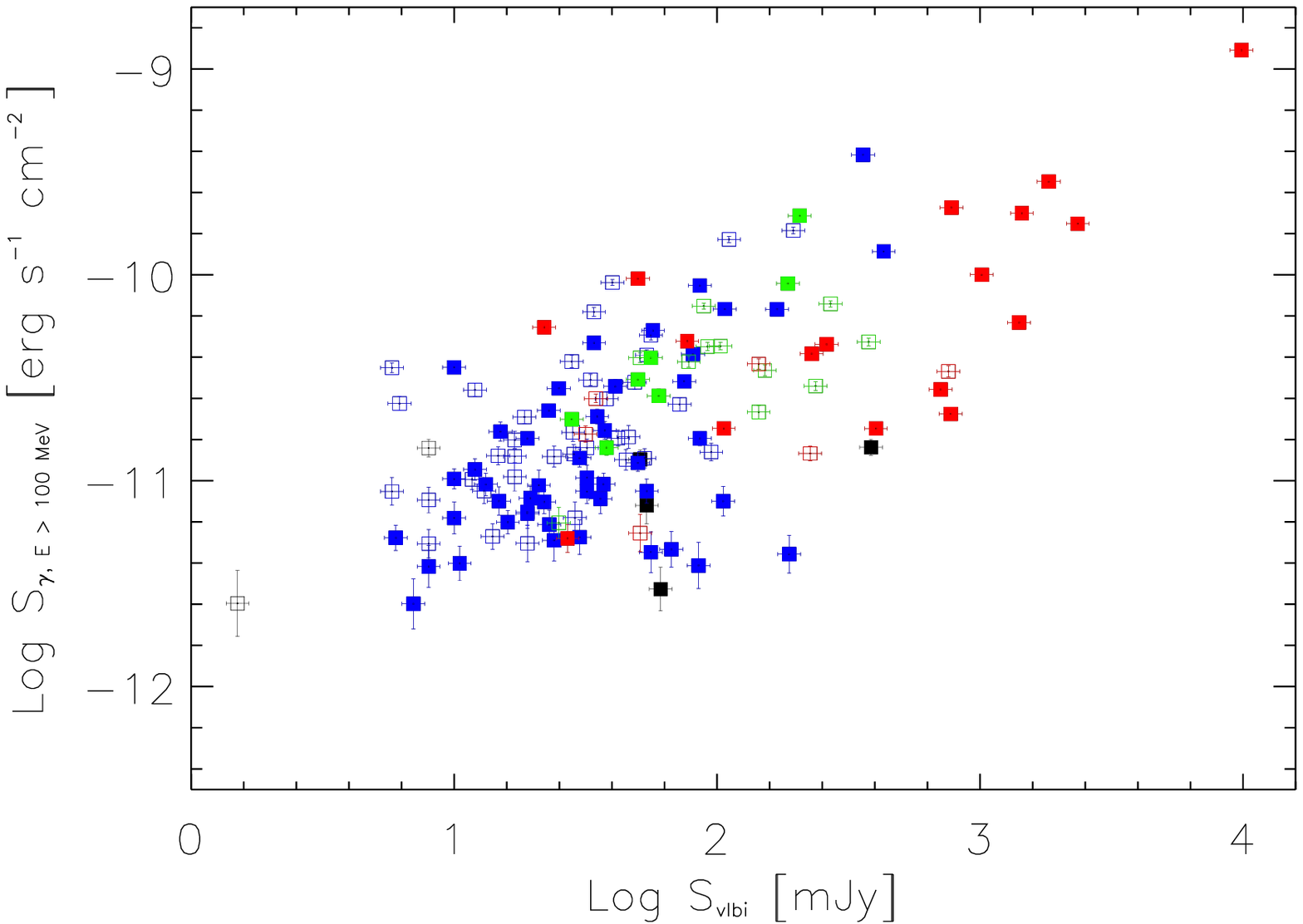}  \\
\end{center}
\caption{\small 1FHL-n sample scatter plots of the VLBI flux density and 1FHL (top left frame) and 3FGL (top right frame) energy flux. 2FHL-n sample scatter plots of the VLBI flux density and 1FHL (bottom left frame) and 3FGL (bottom right frame) energy flux. The different colors represent the HSPs (blue), ISPs (green), LSPs (red), and sources with no spectral classification (black). The sources with and without redshift are indicated by filled and empty symbols, respectively. Images adapted from Lico et al.\ 2017.}
\label{scatter_plots_combo}
\end{figure*}

\begin{table*}[h]
\caption{\small Correlation analysis results for the 1FHL-n and 2FHL-n samples and the various blazar sub-classes.}
\label{tab_corr}     
\centering    
\footnotesize              
\setlength{\tabcolsep}{9pt}
\renewcommand{\arraystretch}{1.3}
\begin{tabular}{lccccc}  
\hline
\hline  
Source type & Catalog & Number of sources & Number of $z$-bins & r-Pearson & Significance \\ 
\hline
\textbf{1FHL-n sample} \\
\hline
All sources & 1FHL & 147 & 14 & -0.05 & $ 0.59 $              \\
            & 3FGL & 147 & 14 &  0.71 & $ <10^{-6} $          \\
BL Lac      & 1FHL & 100 & 9  &  0.12 & $ 0.55 $              \\ 
            & 3FGL & 100 & 9  &  0.70 & $ <10^{-6} $          \\ 
FSRQ        & 1FHL & 44  & 4  & -0.01 & $ 0.99 $              \\  
            & 3FGL & 44  & 4  &  0.49 & $ <10^{-6} $          \\
HSP         & 1FHL & 60  & 5  &  0.57 & $ 1.0\times 10^{-6} $ \\  
            & 3FGL & 60  & 5  &  0.77 & $ <10^{-6} $          \\
ISP         & 1FHL & 23  & 2  &  0.19 & $ 0.40 $              \\ 
            & 3FGL & 23  & 2  &  0.46 & $ 2.5\times 10^{-2} $ \\  
LSP         & 1FHL & 52  & 5  &  0.21 & $ 0.12 $              \\ 
            & 3FGL & 52  & 5  &  0.43 & $ 3.0\times 10^{-6} $ \\ 
\hline
\textbf{2FHL-n sample} \\
\hline 
All sources & 2FHL & 76  & 7  &  0.13 & $ 0.36 $              \\
            & 3FGL & 76  & 7  &  0.72 & $ <10^{-6} $          \\
BL Lac      & 2FHL & 63  & 6  &  0.23 & $ 0.34 $              \\
            & 3FGL & 63  & 6  &  0.73 & $ <10^{-6} $          \\
HSP - with $z$& 2FHL & 48  & 4  &  0.57 & $ 7.0\times 10^{-6} $ \\
              & 3FGL & 48  & 4  &  0.58 & $ <10^{-6} $          \\
HSP - all${\scriptsize ^1}$ & 2FHL & 84  & 8  &  0.61 & $ <10^{-6} $  \\
            & 3FGL & 84  & 8  &  0.53 & $ <10^{-6} $          \\
\hline
\end{tabular}
\begin{threeparttable}
\begin{tablenotes}
\item [1] \footnotesize 2FHL-n HSP sample including sources without known $z$ (see Sect.~\ref{corr_results} for details). 
\end{tablenotes}
\end{threeparttable}
\end{table*}

In Fig.~\ref{scatter_plots_combo} we show the scatter plots between the VLBI flux densities and the 1FHL (top left panel) and 3FGL (top right panel) energy fluxes, for the 1FHL-n sample. Regarding the 2FHL-n sample, the scatter plots between the VLBI flux densities and the 2FHL and 3FGL energy fluxes, are reported in the bottom left and bottom right panels of Fig.~\ref{scatter_plots_combo}, respectively. The different colors represent the HSP (blue), ISP (green) and LSP (red) blazars sub-classes. The sources with no spectral classification are indicated in black color. The sources with and without redshift are indicated by filled and empty symbols, respectively.
The correlation results analysis for both 1FHL-n and 2FHL-n samples are reported in Table~\ref{tab_corr}, where we indicate the number of sources in each subset, the number of redshift bins for the permutations (with >10 objects in each bin), the Pearson correlation coefficient ($r$), and the corresponding statistical significance ($p$).
We investigate the correlation for the full samples, for the optical (FSRQs, and BL Lacs), and spectral (LSPs, ISPs, HSPs) sub-classes. However, in the case of the 2FHL-n only for the BL Lac and HSP blazar sub-classes there are enough sources for obtaining a statistical significant result.

When the 3FGL (0.1 - 300\,GeV) $\gamma$-ray energy fluxes are used, a strong ($r>0.7$) and significant ($p < 10^{-6}$) correlation is found in both 1FHL-n and 2FHL-n samples, as well as in all blazar sub-classes. 
We find an even stronger correlation with respect to the one found in \citep{Ackermann2011}, given that we are using high-resolution VLBI radio observations, which are more representative of the $\gamma$-ray emission zone than low-resolution single dish or interferometic data, which can be contaminated by the emission from extended structures. 
Conversely, the correlation vanishes when the 1FHL (10-500\,GeV) and 2FHL (50\,GeV - 2\,TeV) $\gamma$-ray energies are considered, in both 1FHL-n and 2FHL-n samples and for all of the various blazar sub-classes with the exception of blazars of HSP type.

Given that for assessing the correlation significance we use a method based on the permutations of luminosities, we are using only those sources with known $z$, as a consequence the number of objects in each subset can be reduced and the redshift distribution altered. This effect is particularly important for HSP objects: in the selected samples only about half of them have known $z$. In the case of the 2FHL-n sample, if we include the sources with unknown $z$, by assigning a redshift randomly selected from the sources with known $z$, we obtain an even stronger ($r=0.61$) and significant ($p <10^{-6}$) correlation (see Table~\ref{tab_corr}).

\section{Discussion}

\begin{figure}[t]
\begin{center}
\includegraphics[bb= 84 74 690 550, width= 0.45\textwidth, clip]{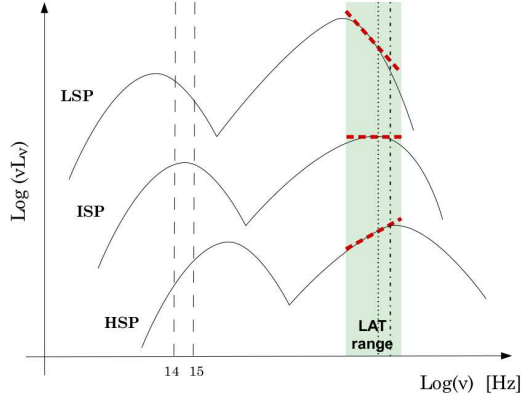} 
\end{center}
\caption{\small Illustrative representation of the blazar SED. LSP, ISP and HSP blazars are represented by the upper, middle and lower curves, respectively. The green area represents the $0.1 - 300$\,GeV 3FGL energy range. The black dotted and dash-dotted lines represent the 1FHL (10\,GeV) and 2FHL (50\,GeV) energy thresholds, respectively. Image from Lico et al.\ 2017.}
\label{schema_SEDs}
\end{figure}

In this work we perform a correlation analysis between radio and $\gamma$-ray emission on two blazar samples extracted from the 1FHL and 2FHL, at energies above 10 and 50\,GeV, respectively. Our main goal is to investigate how the correlation evolves when VHE $\gamma$ rays are considered. While a strong and significant correlation between radio and 0.1-100\,GeV $\gamma$-ray emission was found in various works \citep[e.g.\,,][]{Kovalev2009, Ghirlanda2010, Giroletti2010}, the possible connection between radio and VHE emission is still elusive, mainly because of the lack of an homogeneous coverage of the VHE sky. 
In \citep{Ackermann2011} it was found that the correlation strength depends on the simultaneity of observations in the two observing bands, on the blazar type and on the considered energy band: in particular the correlation strength seems to decrease when higher $\gamma$-ray energies are considered. With respect to the work of \citep{Ackermann2011} we introduce two new ingredients: (1) at radio frequencies we use high-resolution VLBI data and (2) in the $\gamma$-ray band we use data at energies above 10\,GeV. 

From our analysis it emerges that the strong and significant correlation between radio VLBI and 0.1-300\,GeV $\gamma$-ray emission, found for all blazar sub-classes, totally vanishes when $\gamma$-ray energies $>10$\,GeV are considered for all blazar sub-classes with the exception of HSPs.  
An indication of this behavior was found in \citep{Piner2014}, in which the authors found a significant correlation between the parsec-scale radio emission and VHE emission, by using data from the TeVCat\footnote{\url{http://tevcat.uchicago.edu/}.} catalog, for a sample of 41 TeV blazar, mostly of HSP type.

We interpret and explain this correlation behavior within the context of the blazar SED properties. In Fig.~\ref{schema_SEDs} we show an illustrative representation of the blazar SEDs for the various spectral sub-classes: LSPs (upper curve), ISPs (middle curve), and HSPs (lower curve). The green area represent the 0.1-300\,GeV 3FGL energy range, while the dotted and dash-dotted vertical lines represent the 10\,GeV 1FHL and 50\,GeV 2FHL energy thresholds, respectively.

For the most powerful LSP blazars, which have softer spectra with respect to HSPs, the SED high-energy (HE) component peaks at lower energies than those sampled by the LAT. Therefore in our analysis we are considering the decreasing part of the HE spectrum (upper curve in Fig.~\ref{schema_SEDs}), where the $\gamma$-ray flux is quickly decreasing. Moreover, the 1FHL and 2FHL energy ranges are limited to the highest LAT energies (above 10 and 50\,GeV, respectively), where the spectrum has a severe steepening and the emission is strongly dropped, due to cooling losses of the emitting particles \citep[e.g.,][]{Tavecchio2009}.
Conversely, in the case of the less powerful HSP blazars, the energy losses are less severe and the position of the SED high-energy peak is found at higher energies than LSPs and ISPs, in general above $\sim 100$\,GeV. 
In HSPs the part of the HE spectrum in which the cooling losses are dominant is beyond the energy range sampled by the \fermi -LAT \citep{Ghisellini1998}. Therefore the rising part of the HE spectrum is in general sampled both in the 3FGL and 1FHL/2FHL energy ranges.

This sampling effect is in agreement with our results, and can explain why the correlation between radio emission and $\gamma$ rays above 10\,GeV is found only for HSPs. However, an important issue to be taken into account, within this simple scenario, is the variability argument. Blazars are strongly variable sources in all frequency bands, and in case of non-concurrent observations the variability could affect the correlation results. For a detailed discussion on these results see \citep{Lico2017}, where the complete correlation analysis is presented.

\end{document}